\documentclass[aps,prl,twocolumn]{revtex4}
\usepackage[dvipdfm]{graphicx} % Include figure files
\usepackage{bm}% bold math
\usepackage{booktabs}% tabs
\usepackage{amsmath,amsthm,amssymb}
\usepackage{ascmac}
\makeatletter
\AtBeginDocument{\@ifpackageloaded{natbib}{\ifNAT@numbers\if@filesw\immediate\write\@auxout{\string\global\string\NAT@numberstrue}\fi\fi}{}}
\makeatother
\begin{document}

\title{Axial Current driven by Magnetization Dynamics in Weyl Semimetals}
\author{Katsuhisa Taguchi and Yukio Tanaka}

\affiliation{Department of Applied Physics, Nagoya University, Nagoya 464-8603, Japan} %\\ $^2$Department of Physics, Tokyo Institute of Technology, Tokyo, 152-8551, Japan}
%\email{taguchi@rover.nuap.nagoya-u.ac.jp} %taguchi-katsuhisa@ed.tmu.ac.jp
\date{\today}
%                               
% abstract 
\begin {abstract}
We theoretically study the axial current $\bm{j}_5$ (defined as the 
difference between the charge current with opposite chirality) in doped Weyl 
semimetal using a Green's function technique. 
We show that the axial current is 
controlled by the magnetization dynamics 
in a magnetic insulator attached to a Weyl semimetal. 
We find that the induced axial current can be 
detected by using ferromagnetic resonance or the inverse spin Hall effect 
and can be converted into charge current with no accompanying energy loss. 
These properties make Weyl semimetal advantageous for application to low-consumption electronics with new functionality. 
%\\ \\ 
%PACS numbers: MR
%       72.25.-b, %Spin polarized transport (for spin polarized transport devices, see 85.75.-d) 
%       73.43.Qt, %Magnetoresistance 
%       75.47.-m, %Magnetotransport phenomena
%       75.76.+j        Spin transport effects (for devices exploiting spin polarized transport, see 85.75.Hh, 85.75.Mm, and 85.75.Ss)
%------------- DCSE
%PACS numbers: 
%       85.75.-d %Magnetmelectronics;spintronics:device exploiting spin polarized transport or integrated magnetic fields 72.25.-b, %Spin polarized transport (for spin polarized transport devices, see 85.75.-d) 
%       75.47.-m, %Magnetotransport phenomena
%       72.25.-b, %spin polarized transport (for spin polarized transport devices, see 78.75.-d) 
%       [ 75.76.+j      Spin transport effects (for devices exploiting spin polarized transport, see 85.75.Hh, 85.75.Mm, and 85.75.Ss)]
\begin{description}
\item[] \hspace{9.48cm}  PACS numbers: 85.75.-d, 75.47.-m, 72.25.-b %\verb+\pacs{#1}+ command.
%\item[]Keywords: %Ultrafast magnetic vortex switching, Magnetooptical effect, inverse Faraday effect, Spin Berry phase   
\end{description}
\end{abstract}
%\pacs{}
%\keywords{}

%\keywords{}
\maketitle

In spintronics, controlling the propagation of the conduction electron's spin is  a central issue  for wide application of 
low-consumption electronics \cite{rf:Tserkovnyak02,rf:Tserkovnyak03,rf:Kajiwara10,rf:saitoh06,rf:Takahashi08,rf:Kimura07, rf:Ando08}. 
The flow of the spin, i.e., spin current, 
is the difference between the charge current of up-spin and that of 
down-spin and does not accompany any charge current with Joule heating.
This spin current is induced by magnetization dynamics 
at the ferromagnetic metal/normal metal junction \cite{rf:Tserkovnyak02,rf:Tserkovnyak03},
and it can be converted into charge current \cite{rf:Kajiwara10,rf:saitoh06,rf:Takahashi08,rf:Kimura07, rf:Ando08}. 
These properties of spin current 
are useful for low-consumption electricity transmission.

Recently, studies of axial current, which is defined as the difference between the charge current with right-handed 
and that with left-handed fermions, have been revived in the field of quantum chromodynamics \cite{rf:Vilenkin80,rf:Metlitski05,rf:Newman06,rf:Kharzeev07,rf:Kharzeev08,rf:Kharzeev13,rf:Gorbar13}.
A stationary axial current $\bm{j}_5$ exists in the presence of an applied static magnetic field \cite{rf:Vilenkin80,rf:Metlitski05,rf:Newman06,rf:Kharzeev07,rf:Kharzeev08,rf:Kharzeev13,rf:Gorbar13,rf:Chen13,rf:Zyuzin12}. 
This phenomenon is called the chiral separation effect (CSE) \cite{rf:Kharzeev13}. 
Its origin lies in the difference of helicity between right-handed and left-handed fermions.
The helicity $\gamma=\hat{\bm{\sigma}}\cdot\hat{\bm{p}}$ indicates the relative angle between the direction of the spin $\hat{\bm{\sigma}}$ and that of the momentum $\hat{\bm{p}}$ of chiral fermions. 
The helicity of right-handed fermions is $\gamma=+1$, whereas that of left-handed ones is $\gamma=-1$, but both spins are parallel to each other along the applied magnetic field [Fig. 1(a)]. 
Thus, it is remarkable that charge current vanishes in the presence of $\bm{j}_5$ 
only when the numbers of fermions with each chirality are zero \cite{rf:Vilenkin80,rf:Metlitski05,rf:Newman06,rf:Kharzeev07,rf:Kharzeev08,rf:Kharzeev13,rf:Gorbar13}, 
and $\bm{j}_5$ satisfies the conservation law $\dot{\rho}_5 + \bm{\nabla} \cdot \bm{j}_5 = 0$, where $\rho_5$ is the axial charge density.
Recently, focus has been on the detection of the axial current 
and has relied on heavy-ion collision experiments \cite{rf:Kharzeev14}.
\par

It is noted that there is a similarity between the axial current and the spin current. 
Here the axial current transports without accompanying charge current similar to the spin current. 
In fact, the axial current can be decomposed into counterpropagating charge flow with opposite chirality, whose spins are polarized along the applied magnetic field direction. 
Therefore, the axial current is controlled not only by the static magnetic field but also by the magnetization dynamics, which is used to generate spin current in spintronics. 
Moreover, an advantage of using the axial current is its conservative value in contrast to spin current.
One can thus expect new spin transport via the axial current in condensed matter physics. 
Recently, a candidate material hosting Dirac fermions, e.g., Weyl semimetal (WS), has been suggested in condensed matter physics \cite{rf:Wan11,rf:Balents11,rf:Burkov11,rf:Hal12,rf:Xu11,rf:Liu13,rf:Hosur13, rf:tominaga14}. Therefore, studying the transport properties of WS in the context of axial current is of interest.

In this Letter, we study the axial current through a doped WS/magnetic insulator (MI) 
junction [Fig. 1(b)]. Based on a Green's function technique, we derive an analytical formula for the nonequilibrium 
axial current, which is induced by the CSE owing to magnetization dynamics in the MI. 
Such a CSE by magnetization dynamics (DCSE) offers 
the advantage of our being able to control the magnitude of the axial current 
by means of ferromagnetic resonance and is 
useful for detecting the axial current in condensed matter physics. 
Since the present nonequilibrium 
axial current can be transformed into a charge current, 
this axial-current-based electronics, {\it axitronics}, enables applications for low-consumption electricity transmission.
%
%
%%%%%%%%%%%%%% FIG1
\begin{figure}[htbp]\centering \label{fig:1}\ref{fig:1}
\includegraphics[scale=.59]{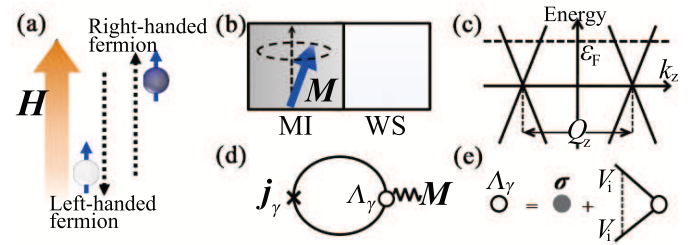} 
\caption{(Color online) 
(a) Schematic illustration of the chiral separation effect. When a magnetic field $\bm{H}$ is applied, right-handed and left-handed fermions are separated along the $\bm{H}$ direction. 
(b) MI/WS junction 
with the dynamical chiral separation effect resulting from the magnetization dynamics $\bm{M}$. 
(c) Schematic illustration of the energy dispersion of 
the WS with time-reversal symmetry breaking and inversion symmetry.
(d) Feynman diagram of charge current from each chiral sector 
in the presence of impurity scattering. 
(e) The vertex function $\bm{\Lambda}_\gamma$ (open circle) and the Pauli matrix (closed circle). 
}\label{fig:fig2}
\end{figure}
%%%%%%%%%%%%%%

The total Hamiltonian we consider is given by 
\begin{align}
\mathcal{H} = \mathcal{H}_{\rm{W}} + \mathcal{H}_{\rm{ex}}  + V_{\rm{i}},
\end{align}
where $\mathcal{H}_{\rm{W}}$, $\mathcal{H}_{\rm{ex}}$ and  $V_{\rm{i}}$
express the Hamiltonian of the conduction electron in doped 
WS, that of exchange coupling between the localized spin in the MI and 
the conduction electron's spin in the WS, and that of impurity scattering in the WS, 
respectively. 
$\mathcal{H}_{\rm{W}}$
is decomposed in each chirality sector as 
$\mathcal{H}_{\rm{W}} = \sum_\gamma \mathcal{H}_{\rm{W},\gamma} $, 
where $\mathcal{H}_{\rm{W},\gamma}$ is given by 
\begin{align}\label{eq:2} %\ref{eq:2}
       \mathcal{H}_{\rm{W}, \gamma} & =  \hbar v_{\rm{F},\gamma} \sum_k \psi^\dagger_{k,\gamma} 
       	[ (\bm{k} - \gamma \bm{Q}/2) \cdot \hat{\bm{\sigma}} ] \psi_{k,\gamma} -\epsilon_{\rm{F}}.
\end{align}
Here $\psi_{k,\gamma} = {}^t\!(\psi_{k,\gamma,\uparrow} \psi_{k, \gamma,\downarrow}), $ and $\psi^\dagger_{k,\gamma}$ are the annihilation and creation operators of the Dirac fermions of each chiral sector $\gamma$, respectively (where indices $\uparrow$ and $\downarrow$ represent spin), $\epsilon_{\rm{F}}$ is the Fermi energy [Fig. 1(c)], and $v_{{\rm{F}},\gamma}=\gamma v_{\rm{F}}$ is the Fermi velocity. 
We assume that a single pair of Dirac cones exists in the WS with 
 inversion-symmetry and  time-reversal-symmetry breaking 
with nonzero $\bm{Q}$. 
The parameter $\bm{Q}$ of Eq. (\ref{eq:2}) denotes  the position of the Weyl node 
with $\gamma \bm{Q}/2$ and its magnitude $|\bm{Q}|$ is the distance between 
two Dirac cones. 
The second term of Eq. (1),
$\mathcal{H}_{\rm{ex}} = \sum_{\gamma=\pm} \mathcal{H}_{\rm{ex},\gamma}$, 
is given by 
\begin{align}
\mathcal{H}_{\rm{ex},\gamma} & = - J_{\rm{ex}} \int d\bm{x} \bm{S} \cdot (\psi^\dagger_\gamma \hat{\bm{\sigma}} \psi_\gamma ), 
\end{align}
where 
$J_{\rm{ex}} >0$ is the exchange coupling constant, %$\frac{1}{2}(\psi^\dagger_\gamma \bm{\sigma} \psi_\gamma )$,  
$\bm{S} = S \bm{n}(\bm{x},t)$ is the classical vector representing the spin structure, $S$ is its magnitude, and $\bm{n}$ is the unit vector representing  the direction, respectively. 
The third term of Eq. (1), $V_{\rm{i}}$, represents nonmagnetic impurity scattering, which 
causes a relaxation time $\tau$ of the transport of conduction electrons in the WS. 
%The relaxation time is evaluated from the random impurity averages. 

In the following calculation, 
$\bm{Q}$ is chosen to be parallel to the quantization axis of the localized spin ($z$ axis) as $\bm{Q}= Q_z \bm{z}$ and $Q_z$ is a constant that is independent of time. 
In addition, we incorporate the term proportional to $\bm{Q}$ in $H_{{\rm{W}},\gamma}$ into $\mathcal{H}_{\rm{ex},\gamma}$ 
by using the following transformation: $\bm{S}  \to \bm{S}' =(S_x, S_y, S_z - \frac{\hbar v_{\rm{F}}}{2J_{\rm{ex}}}Q_z )$.
This transformation enables us to calculate the axial current rather easily. 
Then, we assume that the effect of 
$\mathcal{H}_{\rm{ex},\gamma}$ is weak and can be treated as a perturbation. 
This  condition is satisfied by $J_{\rm{ex}} |\bm{S}'| \tau/\hbar\ll 1 $  
within the  diffusive transport regime.

To consider the axial current created by the  DCSE, we will calculate the current $\bm{j}_\gamma$  using the above assumptions. 
We define the charge current of each chirality sector $\gamma$ as $\bm{j}_\gamma = - ev_{\rm{F},\gamma} \langle \psi^\dagger_\gamma \bm{\sigma} \psi_\gamma \rangle $ from the conservation law $\dot{\rho}_\gamma = - \bm{\nabla}\cdot \bm{j}_\gamma$, where $\rho_\gamma \equiv -e \langle \psi^\dagger_\gamma \psi_\gamma \rangle $ is 
the charge density of chirality $\gamma$. 
The current is represented by using the same space and time of lesser Green's functions $G^<_\gamma  =  \langle \psi^\dagger_\gamma \psi_\gamma \rangle /(-i\hbar)$ as
\begin{align} \label{eq:4} %\ref{eq:4}
j_{i,\gamma} (\bm{x}, t) 
        & =  i\hbar  ev_{\rm{F},\gamma}   {\rm{tr}}[\hat{\sigma} _i G_\gamma^<(\bm{x}, t:\bm{x}, t )].
\end{align}
By using the Fourier transformation, 
the Dyson equation of $G_\gamma^<$ is given 
by    
\begin{align}\notag 
G^<_{\bm{k},\bm{k'}, \omega,\omega',\gamma} & = g^<_{\bm{k},\omega,\gamma} \delta_{\bm{k},\bm{k'}} \delta_{\omega,\omega'}
        \\  \label{eq:5} %\ref{eq:5}
         - \frac{J_{\rm{ex}}}{V} &\sum_{\bm{q},\Omega} [ g_{\bm{k},\omega,\gamma} 
                \hat{\bm{\sigma}} \cdot  \bm{S}'_{\bm{q},\Omega} G_{ \bm{k}+\bm{q}, \bm{k'} \omega+\Omega,\omega',\gamma} ]^<, 
\end{align}
where $V$ is the system volume and $g^<_{\bm{k},\omega,\gamma} $ is the Green's function of $\mathcal{H}_{{\rm{W}},\gamma}$ including $V_{\rm{i}}$, 
\begin{align}\label{eq:6} %\ref{eq:6}
g^r_{\bm{k},\omega,\gamma} 
        & = \left[ \hbar \omega + \epsilon_{\rm{F}} -  \hbar v_{{\rm{F}},\gamma} \bm{k}\cdot \hat{\bm{\sigma}} + i\eta \right]^{-1}.
\end{align}
Here $g^r (g^a)$ is the retarded (advanced) Green's function. 
$\eta \equiv \hbar/(2\tau) = n_i u_i^2 \nu_e/4$ %\frac{\hbar}{2\tau} = \frac{n_i u_i^2 \nu_e}{4}$
 is the self-energy of $V_{\rm{i}}$, 
 where $n_i$, $u_i$, and $\nu_{e}$ are the concentration of impurities, 
 the potential energy of impurities, 
 and the density of states at $\epsilon_{\rm{F}}$, respectively. 
$j_{\mu,\gamma} $ is diagrammatically represented in Fig. 1(d) and is obtained from Eqs. (\ref{eq:4})--(\ref{eq:6}) as
\begin{align} \label{eq:spin density-first1} %\ref{eq:spin density-first1}
j_{i,\gamma}  & =  \frac{-i \hbar J_{\rm{ex}} ev_{\rm{F},\gamma} }{V}\sum_{\bm{q},\Omega}e^{-i(\bm{q}\cdot\bm{x}-\Omega t)} \Pi_{ij,\gamma} (\bm{q},\Omega)S'^j_{\bm{q},\Omega} , 
\\
\Pi_{ij ,\gamma} 
        & = \frac{1}{V} \sum_{\bm{k}, \omega} {\rm{tr}} [ \hat{\sigma}_i g_{\bm{k}-\frac{\bm{q}}{2},\omega-\frac{\Omega}{2},\gamma} \hat{\Lambda}_{j,\gamma}  g_{\bm{k}+\frac{\bm{q}}{2},\omega+\frac{\Omega}{2},\gamma} ]^<,
\end{align}
where $\Pi_{ij,\gamma} $ is the spin-spin correlation function and $\hat{\Lambda}_{j,\gamma}$ is the vertex function of $V_{\rm{i}}$ expressed in Fig. 1(e). The vertex function is given by
\begin{align} \notag 
\hat{\Lambda}_{\mu,\gamma} & \equiv  \sum_{n=0}^\infty \prod_{\nu=0}^n \sum_{k_1 \cdots k_n }(n_iu_i^2)^\nu  ( g_{{\bm{k_\nu}-\frac{\bm{q}}{2}}, \omega-\frac{\Omega}{2},\gamma})^\nu \hat{\sigma}_\mu ( g_{{\bm{k_\nu} + \frac{\bm{q}}{2}}, \omega+\frac{\Omega}{2},\gamma})^\nu 
\\
 & = [1- \hat{\Gamma}_{\mu,\gamma} ]^{-1} = \Lambda_{\mu \zeta,\gamma} \hat{\sigma}_\zeta, 
 \\
 \hat{\Gamma}_{\mu,\gamma} & = \frac{1}{V} \sum_{\bm{k}} n_iu_i^2  g_{\bm{k}-\frac{\bm{q}}{2}, \omega-\frac{\Omega}{2},\gamma}  \hat{\sigma}_\mu  g_{\bm{k}+\frac{\bm{q}}{2}, \omega+\frac{\Omega}{2},\gamma} = \Gamma_{\mu \zeta,\gamma} \hat{\sigma}_\zeta, 
\end{align}
where $\Lambda_{\mu \zeta,\gamma}$ and $\Gamma_{\mu \zeta,\gamma}$ are $4\times4$ matrices with indices $\mu, \zeta = 0, x, y, z$. 
We calculate $\Pi_{ij,\gamma} $ by using $g^<_{\bm{k},\omega,\gamma} =f_\omega (g^a_{\bm{k},\omega,\gamma} - g^r_{\bm{k},\omega,\gamma}) $ \cite{rf:book1}, where $f_\omega$ is the Fermi distribution function. 
Now, we only consider the nonequilibrium component of 
$j_{i,\gamma}$ \cite{rf:note0}. 
The dominant contribution is obtained by using $\frac{\hbar}{\epsilon_{\rm{F}}\tau} \ll1$, expanding with $q/k_{\rm{F}} \ll1 $ and $\Omega \tau\ll1$, and assuming isotropic $\bm{q}$ as 
\begin{align} \notag 
\Pi_{ij,\gamma}%(\bm{q},\Omega) 
= & \sum_{\bm{k}, \omega}  (f_{\omega+\frac{\Omega}{2}} -f_{\omega-\frac{\Omega}{2}} ) {\rm{tr}} [  \hat{\sigma}_i g^r_{\bm{k}-\frac{\bm{q}}{2},\omega-\frac{\Omega}{2},\gamma} \hat{\Lambda }_{j,\gamma} g^a_{\bm{k}+\frac{\bm{q}}{2},\omega+\frac{\Omega}{2},\gamma}]  %+ o\left( \frac{\hbar}{\epsilon_{\rm{F}}\tau}\right)
\\   \label{eq:11} %\ref{eq:11}
 = & - \frac{ \nu_e \Omega \tau }{2\hbar}  \left[  \delta_{ij} - \frac{\frac{3}{2}D_\gamma q_iq_j}{\frac{3}{2}D_\gamma q^2 +i\Omega } \right],
\end{align}
where $D_\gamma= \frac{1}{3} v_{\rm{F},\gamma}^2 \tau = \frac{1}{3} v_{\rm{F}}^2 \tau = D $ is the diffusion constant. Here $\Pi_{ij,\gamma}$ gives
\begin{align}  \label{eq:12} %\ref{eq:12}
        j_{i, \gamma}
                & =  \frac{e v_{\rm{F},\gamma}  J_{\rm{ex}} \nu_e}{2V}  \sum_{\bm{q},\Omega}e^{i(\Omega t-\bm{q}\cdot\bm{x})} 
         i\Omega\tau \left[    \delta_{ij}
                -\frac{ \frac{3}{2}D_\gamma q_i q_j }{ \frac{3}{2} D_\gamma q^2 + i\Omega} \right]   S^j_{\bm{q},\Omega}.
\end{align}
The second term in the above equation is determined by the charge density resulting from the magnetization dynamics. 
$\rho_\gamma$ is calculated by using $\Pi_{0j,\gamma}$ and is expressed by 
\begin{align}
\rho_\gamma 
        \label{eq:13 charge} %\ref{eq:13 charge}
                & =  \frac{-e v_{\rm{F},\gamma} J_{\rm{ex}} \nu_e}{2V}\sum_{\bm{q}, \Omega}e^{i(\Omega t - \bm{q}\cdot\bm{x})}     
                        \frac{\Omega \tau q_j  }{ \frac{3}{2}D_\gamma q^2 + i\Omega  } S^j_{\bm{q},\Omega}
                        \\ \label{eq:14 charge} %\ref{eq:14 charge}
        & =  - \frac{1}{2 }e v_{\rm{F},\gamma} J_{\rm{ex}} \nu_e \tau    \bm{\nabla} \cdot  \partial_t \langle \bm{S}\rangle_{\rm{D}} ,
\end{align}
where $\langle \bm{S}\rangle_{\rm{D}}$ is defined by the convolution of  $\bm{S}$ and a diffusive propagation function $\mathcal{D}$ \cite{rf:note1} given by   \begin{align} \label{eq:15} %\ref{eq:15} 
\langle \bm{S} \rangle_{{\rm{D}}}
        & \equiv \int_{-\infty}^\infty dt' \int d\bm{x'}  \mathcal{D}(\bm{x}-\bm{x'}, t-t') \bm{S}( \bm{x'},t' ),
        \\ \label{eq:16} %\ref{eq:16}
        \mathcal{D} (\bm{x}, t) & \equiv \frac{1}{V} \sum_{\bm{q},\Omega} e^{-i(\bm{q}\cdot\bm{x}-\Omega t)} \frac{1}{\frac{3}{2}D_\gamma q^2 + i\Omega}.
\end{align}
Therefore, we obtain the current  
\begin{align}  \label{eq:17 current} %\ref{eq:17 current}
\bm{j}_\gamma & = \frac{e v_{\rm{F},\gamma} J_{\rm{ex}} \nu_e \tau}{2} \dot{\bm{S}} - \frac{3D_\gamma}{2} \bm{\nabla}  \rho_\gamma.
\end{align}
It is noted that, from Eqs. (\ref{eq:12}) and (\ref{eq:13 charge}), $\rho_\gamma$ and $\bm{j}_\gamma$ satisfy the conservation law $\dot{\rho}_\gamma +  \bm{\nabla}\cdot \bm{j}_\gamma =0$.

{\it Axial current.}---Now, we turn to a discussion of the charge current and the charge density after the summation over the index of the chirality $\gamma$.
Since $\bm{j}_\gamma$ is proportional to the chirality from Eq. (\ref{eq:12}), the directions of $\bm{j}_+$ and $\bm{j}_-$ are opposite to each other. 
In the same way, $\rho_+$ becomes $\rho_+ = -\rho_-$ from Eq. (\ref{eq:14 charge}). 
%In the same way, $\rho_+$ and $\rho_-$ have opposite sign each other from Eq. (\ref{eq:14 charge}). 
Thus, the total charge current and density vanish:
\begin{align} \label{eq:18} %\ref{eq:18}
        \begin{matrix}
                \bm{j}_+ +  \bm{j}_-  =0, \\
                \rho_+ +  \rho_-  =0.
        \end{matrix}
\end{align}
However, from Eqs. (\ref{eq:14 charge}) and (\ref{eq:17 current}), the axial current $\bm{j}_{5} \equiv \bm{j}_+ -  \bm{j}_-$ and the axial charge $\rho_{5} \equiv \rho_+ -  \rho_-$ are given by   
\begin{align} \label{eq:19} %\ref{eq:19}
\bm{j}_5
        & = ev_{\rm{F}} J_{\rm{ex}} \nu_e \tau \dot{\bm{S}} - \frac{3}{2}D  \bm{\nabla} \rho_{5}, 
        \\  \label{eq:19-2} %\ref{eq:19-2}
\rho_{5}
        & =  -ev_{\rm{F}} J_{\rm{ex}} \nu_e \tau   \bm{\nabla} \cdot  \partial_t \langle \bm{S}\rangle_{\rm{D}}.
\end{align} 
This $\bm{j}_{5}$ is triggered by the DCSE.
We can decompose $\bm{j}_{5}$ 
into a local component 
$\bm{j}_5^{\rm{L}}$ and 
a nonlocal one $\bm{j}_5^{\rm{N}}$ with 
$\bm{j}_5 \equiv \bm{j}_5^{\rm{L}} + \bm{j}_5^{\rm{N}}$. 
The first term of Eq. (\ref{eq:19}) corresponds to 
$\bm{j}_5^{\rm{L}} $ parallel to $\dot{\bm{S}}$ and is induced by the time-dependent 
magnetization dynamics $\dot{\bm{S}}$.
The second term of Eq. (\ref{eq:19}) expresses 
$\bm{j}_5^{\rm{N}}$, which is driven by the spatial gradient of the axial charge $\rho_{5}$ and is parallel to its gradient. 
Here, $\rho_{5}$ is triggered by the time and 
spatial dependence of the magnetization dynamics, $\bm{\nabla} \cdot  \partial_t  \langle \bm{S}\rangle_{\rm{D}}$ \cite{rf:difference between spin current and axial current}. 
Here $\langle \bm{S} \rangle_{\rm{D}}$ expresses the 
diffusion propagation by  random impurity scattering.
From Eq. (\ref{eq:19-2}), the DCSE triggers only the nonlocal component of the axial charge.

We will compare Eq. (17) with 
the charge current and the spin generation resulting from 
the magnetization dynamics 
at the junction of a MI deposited on the surface of a topological insulator (TI). 
Then, the charge current stemming from each chirality 
owing to magnetization dynamics becomes $\bm{j}_\gamma \propto v_{\rm{F},\gamma} \dot{\bm{S}}$ \cite{rf:Qi08,rf:nomura10, rf:ueda12}. 
This current is proportional to each chirality. 
Although $\bm{j}_\gamma$ is proportional to $\gamma$ 
similarly to that in Eq. (\ref{eq:17 current}),
there is no summation of helicity index $\gamma$ on the 
surface of the TI that is different from that of the WS. 
On one side of the surface of the TI, only the Dirac cone with 
 right- or left-handed chirality exists, whereas, in the bulk of the WS, 
there are Dirac cones with both chiralities \cite{rf:Nielsen}.  
%%
%%%%%%%%%%%%%% FIG2-1-b
\begin{figure}[tbp]\centering \label{fig:2}\ref{fig:2}
\includegraphics[scale=.32]{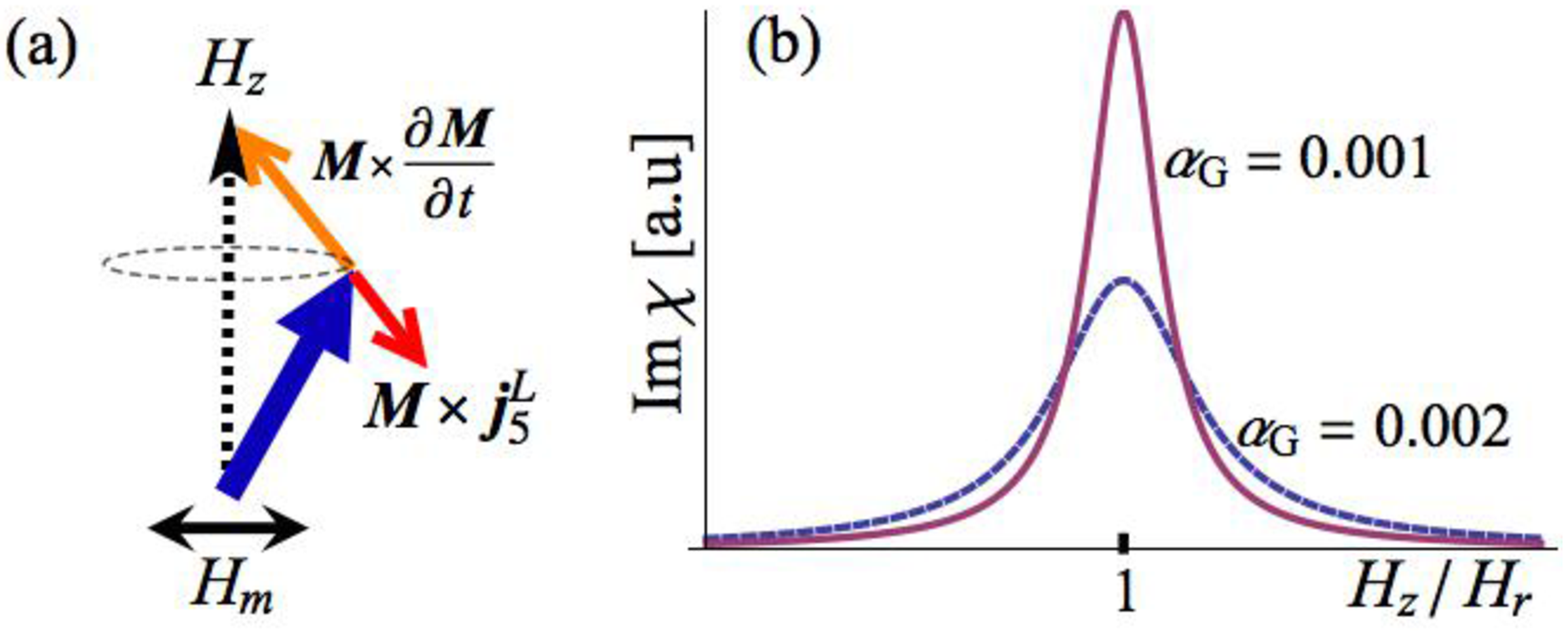} 
\caption{(Color online) 
(a) Magnetic precessional motion after the generation of the axial current 
$\bm{j}_{5}^{\rm{L}}$ in the presence of the applied magnetic field 
$H_z || \bm{z}$ and ac magnetic field $H_m \perp \bm{z}$. 
The local axial current $\bm{j}_5^{\rm{L}}$ triggers spin torque ($\propto \bm{M}\times \bm{j}_5^{\rm{L}}$), which prevents the damping ($\propto \bm{M}\times \partial_t \bm{M}$). (b) The induced torque can be detected from the half-width value of the permeability $\chi$ depending on $H_z/H_r$ at the resonance frequency, where $H_r$ is the resonance magnetic field and $\alpha_{\rm{G}}$ is the Gilbert constant. }\label{fig:fig2}
\end{figure}
%%%%%%%%%%%%%%
%%

{\it Detection of $\bm{j}_{5}^{\rm{L}}.$}---First, we consider the magnetization dynamics after the generation of $\bm{j}_5$. 
 $\bm{j}_5$ can be interpreted as the total spin $\bm{s} = \bm{j}_5/(-2ev_{\rm{F}})$ in the WS, because of  spin-momentum locking. %%
Therefore, $\bm{j}_5$ like $\bm{s}$ plays the role of an exchange field acting on the magnetization. 
The exchange field $\bm{b}\equiv -\frac{1}{\hbar g \mu_B  }\frac{\delta{\mathcal{H}}}{\delta \bm{S}}$ is given by 
\begin{align} \label{eq:20} %\ref{eq:20}
\bm{b} 
        &  =  -  \frac{J_{\rm{ex}}}{ 2 e g \mu_B  v_{\rm{F}} }  \bm{j}_5.
\end{align}
The magnetization dynamics caused by $\bm{b}$ is obtained 
from the Landau--Lifshitz--Gilbert equation \cite{rf:Chikazumi,rf:Tatara08},
which is given by  
\begin{align} \label{eq:21} %\ref{eq:21}
\dot{\bm{M}} & =  \frac{g \mu_B}{\hbar} \mu  \bm{H} \times \bm{M} + \frac{\alpha_{\rm{G}}}{M} \bm{M} \times \dot{\bm{M}} + \bm{\mathcal{T}}_{\rm{e}}, 
\end{align} 
where $\bm{M}=- g \mu_B \bm{S}/a^3$ is the magnetization, $g$ is the Land\'{e} factor, $\mu_B$ is the Bohr magneton, $a$ is the lattice constant, $\mu$ is  permeability,  $\bm{H}$ is the applied magnetic field, $\alpha_{\rm{G}}$ is Gilbert damping representing relaxation of the magnetization dynamics, and $\bm{\mathcal{T}}_{\rm{e}}$ is the torque of conduction electron spin $\bm{s}$, the so-called spin torque \cite{rf:Tatara08}. 
From Eqs. (19) and  (20), this torque $\bm{\mathcal{T}}_{\rm{e}} = \frac{g \mu_B}{\hbar}  \bm{b} \times \bm{M} = \frac{J_{\rm{ex}} }{2e v_{\rm{F} } } (\bm{M} \times \bm{j}_{5}^{\rm{L}} + \bm{M} \times \bm{j}_{5}^{\rm{N}} )$ is given by
\begin{align} \label{eq:22} %\ref{eq:22}
\bm{\mathcal{T}}_{\rm{e}}
        & = - \frac{J^2_{\rm{ex}} \nu_e  \tau S}{\hbar M }    \biggl[ \bm{M} \times \dot{\bm{M}}  + \frac{3}{2} D \bm{M} \times \bm{\nabla} \left( \bm{\nabla} \cdot \partial_t \langle \bm{M} \rangle_{\rm{D}} \right) \biggr]        .
\end{align}
The first term of Eq. (\ref{eq:22}) corresponds to $\bm{M} \times \bm{j}_{5}^{\rm{L}}$ and shows that the torque due to $\bm{M} \times \bm{j}_{5}^{\rm{L}}$ suppresses the relaxation of the magnetization dynamics [Fig. 2(a)] from Eq. (22). 
The second term of Eq.  (\ref{eq:22}) is caused by $\bm{M} \times \bm{j}_{5}^{\rm{N}}$;
 its direction is perpendicular to $\bm{M}$ and $\bm{\nabla} [ \bm{\nabla} \cdot \partial_t  \langle \bm{M} \rangle_{\rm{D}} ]$, which depends on the magnetic structure. 
From Eqs. (\ref{eq:21}) and (\ref{eq:22}), 
the torque $\bm{M} \times \bm{j}_5^{\rm{L}}$ can be detected by using magnetic resonance before and after the generation of $ \bm{j}_{5}^{\rm{L}}$, since the damping coefficient $\alpha_{\rm{G}}$ is experimentally estimated from the half-width value of the permeability at magnetic resonance \cite{rf:Chikazumi,rf:Mizukami02}.
For example, we simply apply an external magnetic field $\bm{H} = H_z \bm{z}$ 
and an ac magnetic field $\bm{H}_m \perp \bm{z}$ at the resonance frequency $\omega_0$ in the MI, 
whose $\bm{M}$ is spatially uniform as shown in Fig. 2(a). 
Then,  $ \bm{j}_{5}^{\rm{N}} \propto \bm{\nabla} [ \bm{\nabla} \cdot \partial_t  \langle \bm{M} \rangle_{\rm{D}}]$ is zero\cite{rf:spin current does not affect the damping}
and  $\bm{j}_{5} = \bm{j}_{5}^{\rm{L}}$ is induced at the interface between the MI and the WS. 
As a result, when $H_z$ is equal to the resonant magnetic field $H_r$,  the half-width value $\Delta(H_z)$ [Fig. 2(b)] becomes  
\begin{align}\label{eq:23} %\ref{eq:23}
\Delta(H_z/H_r=1) 
        & = 2\omega_0  \bigl( \alpha_{\rm{G}} - J^2_{\rm{ex}} \nu_e \tau S / \hbar \bigr).
\end{align}
This equation means that before and after the generation of $ \bm{j}_{5}^{\rm{L}}$, the half-width value changes from $2\omega_0\alpha_{\rm{G}} $ by $2\omega_0J^2_{\rm{ex}} \nu_e \tau S / \hbar$, which is caused by the presence of $\bm{j}_5^{\rm{L}}$ from Eqs. (22) and (23). 
When we chose the parameters 
$J_{\rm{ex}}/\epsilon_{\rm{F}} =0.01$, $\tau=6\times 10^{-14}$ s, $\epsilon_{\rm{F}} \nu_e =1$, and $S=5/2$,  the change in damping is estimated as $ J^2_{\rm{ex}} \nu_e \tau S /\hbar \sim 2 \times 10^{-3}$. The order of $\alpha_{\rm{G}}$ is reported as $10^{-3}$ in magnetic metals \cite{rf:Mizukami02} and $10^{-5}$ in MIs \cite{rf:Kajiwara10}. Therefore, the change of the half-width value should be measurable by $\Delta(H_z/H_r)$.

%%%%%%%%%%%%%% FIG3
\begin{figure}[tbp]\centering \label{fig:2}\ref{fig:2}
\includegraphics[scale=.16]{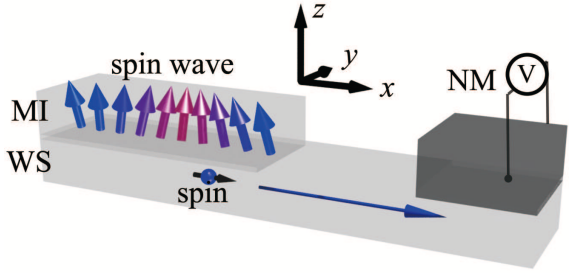} 
\caption{(Color online) 
Geometry for detections of the nonlocal axial current $\bm{j}_{5}^{\rm{N}}$ 
at the MI/WS/NM junction. $\bm{j}_{5}^{\rm{N}}$ is triggered by  time-dependent magnetization dynamics, such as a spin wave propagating along the $x$ axis. 
This $\bm{j}_{5}^{\rm{N}} || \bm{x}$, which is interpreted as spin $\bm{s}^{\rm{N}} || \bm{x}$, propagates 
isotropically and accumulates at the edge of the WS.
}\label{fig:fig2}
\end{figure}
%%%%%%%%%%%%%%

{\it Detection of $\bm{j}_{5}^{\rm{N}}.$}---Next, we discuss an experimental 
method for the 
detection of the nonlocal part of the axial current $\bm{j}_{5}^{\rm{N}}$ from the diffusion equation. 
We do not consider the contribution from $\bm{j}_{5}^{\rm{L}}$. 
The diffusion equation is given by Eqs. (15), (16), and  (19) as \cite{rf:note2} 
\begin{align} \label{eq:25 diffusion} %\ref{eq:25 diffusion}
(\partial_t - \frac{3}{2}D\nabla^2) \bm{j}_{5}^{\rm{N}} = -\frac{3e v_{\rm{F}}  J_{\rm{ex}} \nu_e a^3}{2 g \mu_{\rm{B}} } D \bm{\nabla} (\bm{\nabla} \cdot   \dot{\bm{M}}).
\end{align}
This equation shows that $\bm{j}_{5}^{\rm{N}}$ produced by the source 
term  $D \bm{\nabla} (\bm{\nabla} \cdot   \dot{\bm{M}})$ 
isotropically propagates. 
%%%%%%%%%%%%%%%%%%%%%%%%%
This $\bm{j}_{5}^{\rm{N}}$ can be interpreted as the conduction electron's 
spin $\bm{s}^{\rm{{N}}}$ with $\bm{s}^{\rm{{N}}} =\bm{j}^{\rm{{N}}}_5/(-2ev_{\rm{F}})$. 
Then, Eq. (25) is regarded as a diffusive equation with spin. 
From this equation, we find that $\bm{s}^{\rm{N}} \propto \bm{j}_{5}^{\rm{N}}$ accumulates at the edge 
of the sample and its accumulation 
can be electrically detected at the MI/WS/normal metal (NM) junction (Fig. 3) by using the method established in spintronics \cite{rf:Kajiwara10,rf:saitoh06,rf:Takahashi08,rf:Kimura07,rf:Ando08}. 
For example, we assume that $\bm{M}$ in the MI has a  
spatial dependence only along the $x$ axis and 
that the NM has a spin-orbit interaction. 
Then, 
$\bm{\nabla} (\bm{\nabla} \cdot   \dot{\bm{M}})$
is parallel to  the $x$ axis   
and triggers $\bm{s}^{\rm{N}}|| \bm{x} $. 
The induced spin  $\bm{s}^{\rm{N}} || \bm{x} $ is isotropically propagating and accumulating at the edge of the WS. 
The accumulated spin can be sinked into the NM along the $z$ axis \cite{rf:Kajiwara10,rf:saitoh06,rf:Takahashi08,rf:Kimura07,rf:Ando08} 
(flow of spin $\bm{I}_s || \bm{z}$ and $\bm{s}^{\rm{N}} || \bm{x} $) and is converted into charge current $\bm{j} \propto \bm{s}^{\rm{N}} \times \bm{I}_s$ parallel to the $-y$ axis through the inverse spin Hall effect \cite{rf:Kajiwara10,rf:saitoh06,rf:Takahashi08,rf:Kimura07,rf:Ando08}.
We notice that $\bm{j}_5^{\rm{N}}$ propagates without any accompanying 
charge current [see Eq. (18)] and functions similarly to the spin current \cite{rf:note3}. 
However, in contrast to spin current, the axial current is a conservative quantity. 
Thus, we expect that  $\bm{j}_5^{\rm{N}}$
is useful for detection of the axial current electrically and for application to low-consumption electricity transmission.

{\it Gauge invariance.}---We find that $\bm{j}_5$ and $\rho_5$ are proportional to $\dot{\bm{S}}$ from Eqs. (\ref{eq:19}) and  (\ref{eq:19-2}) 
because of the gauge invariance in the WS. 
Owing to  spin-momentum locking, 
$\bm{S}$ plays a role like the electromagnetic vector potential as 
$\mathcal{H}_{{\rm{W}},\gamma}+ \mathcal{H}_{\rm{ex},\gamma} \propto \bm{\sigma} \cdot (\bm{k}-\frac{e}{\hbar} \bm{\mathcal{A}}_\gamma)$, 
where the vector potential $\bm{\mathcal{A}}_\gamma = J_{\rm{ex}} \bm{S}/(ev_{\rm{F},\gamma})$ is conjugate to $\bm{j}_\gamma$. 
Therefore, the observable quantity should 
be proportional to the gauge invariant form as $-\partial_t \bm{\mathcal{A}}_\gamma \equiv \bm{\mathcal{E}}_\gamma$  
 or $\bm{\nabla} \times \bm{\mathcal{A}}_\gamma \equiv \bm{\mathcal{B}}_\gamma$.
 The axial current and 
charge are induced by an effective electric field $\bm{\mathcal{E}}_\gamma$ and $ \bm{\nabla}\cdot \langle \bm{\mathcal{E}_\gamma}\rangle_{\rm{D}}$, 
respectively,  as shown from 
Eqs. (\ref{eq:19}) and (\ref{eq:19-2}).  

In conclusion, we studied the nonequilibrium axial current density $\bm{j}_5$ and axial charge density $\rho_5$ based on a Green's function technique 
at the MI/doped WS junction. 
We find that the DCSE drives the axial current by time-dependent magnetization dynamics, 
$\dot{\bm{S}}$, as expected from the gauge invariance of $\bm{S}$. 
The axial current can be decomposed into local and nonlocal ones. 
Based on our results, we discuss a  procedure for the detection of the local and nonlocal axial current  
by using magnetic resonance and the inverse spin Hall effect, respectively.  
The DCSE induces $\bm{j}_5$ with no accompanying charge transport, and $\bm{j}_5$ can be converted into charge current at the MI/WS/NM junction.
These properties of $\bm{j}_5$ can be useful for the 
application of WS to low-consumption electronics. 
Thus, the present letter has explored a new area of axial-current-based electronics, {{\it axitronics}}.

\begin{acknowledgments}
This work was supported by Grants-in-Aid for Young Scientists (B) (No. 22740222 and No. 23740236) and by  Grants-in-Aid for Scientific Research on Innovative Areas ``Topological Quantum Phenomena'' (No. 22103005 and No. 25103709) from the Ministry of Education, Culture, Sports, Science, and Technology, Japan (MEXT). K.T. acknowledges support from the JSPS.
\end{acknowledgments}


\begin{references}
\bibitem{rf:note0}We ignore contributions from $\Pi_{ij,\gamma}(\bm{q}, \Omega=0)$, which implies an equilibrium quantity. Since $\bm{Q}$ is independent of time, $\bm{Q}$ does not contribute to  $\Pi_{ij,\gamma}(\bm{q}, \Omega\neq0)$.
\bibitem{rf:difference between spin current and axial current}
We find that $\rho_5$ is proportional to the spin current $j_{{\rm{s}},i}^\alpha$ in the WS as 
$j_{{\rm{s}},i}^\alpha \propto v_{\rm{F}} \rho_5 \delta_{i \alpha}$, 
where the spin current is defined by $\dot{s}^\alpha + \nabla_i j_{{\rm{s}},i}^\alpha = \mathcal{T}^\alpha$ 
and $\mathcal{T}^\alpha$ is a spin relaxation.
\bibitem{rf:spin current does not affect the damping}
When the magnetization is spatially uniform, the spin current $j_{{\rm{s}},i}^\alpha \propto \rho_5 \delta_{i \alpha}$ also zero, and the spin current does not contribute to the change of the damping of the magnetization dynamics.
\bibitem{rf:note2}The diffusion equation is given by acting the differential equation of $\mathcal{D}$ onto $\bm{j}_{5}^{\rm{N}}$ from the left side, where the differential equation is defined from Eq. (16) as $(\partial_t - \frac{3}{2}D\nabla^2) \mathcal{D}(\bm{x}-\bm{x}', t-t') = \delta(t-t')\delta^3(\bm{x}-\bm{x}')$ and Eq. (\ref{eq:19}).
\bibitem{rf:note3}From Eq. (25), $s^{{\rm{{N}}},i}$ due to $j_5^{{\rm{N}},i}$ has no damping because $s^{{\rm{{N}}},i}$ follows the conservative law, $ \partial_t {s}^{{\rm{{N}}},i}  + \partial_a j_{\rm{{s}},a}^i=0$,  where $j_{\rm{{s}},a}^i = -\frac{3}{2}D ( \partial_a s^{{\rm{{N}}},i} + \frac{J_{\rm{ex}} \nu_e a^3 }{2 g \mu_{\rm{B}}}  \partial_i M^a   )$.
\end{references}
\end{document}